\documentclass[journal,10pt]{IEEEtran}

\IEEEoverridecommandlockouts
\usepackage{cite}
\usepackage{amsmath,amssymb,amsfonts}
\usepackage{algorithmic}
\usepackage{graphicx}
\usepackage{textcomp}
\usepackage{xcolor}
\def\BibTeX{{\rm B\kern-.05em{\sc i\kern-.025em b}\kern-.08em
    T\kern-.1667em\lower.7ex\hbox{E}\kern-.125emX}}

\usepackage{color}
\usepackage{amsmath,bm}
\definecolor{MyDarkBlue}{rgb}{0,0.08,0.45}
\definecolor{yellow}{rgb}{0.99,0.99,0.70}
\definecolor{myback}{RGB}{204,232,207}  
\definecolor{white}{rgb}{1.0,1.0,1.0}
\usepackage{subeqnarray}
\usepackage{cases}   
\definecolor{black}{rgb}{0.00,0.00,0.00}
\usepackage[scriptsize]{subfigure}
\usepackage{mathtools}
\usepackage{bm}
\usepackage{makecell}

\usepackage{cite}
\usepackage{verbatim}
\usepackage{epsfig,bbm}
\usepackage[colorlinks,bookmarksopen,bookmarksnumbered,citecolor=red,urlcolor=red,]{hyperref}
\usepackage{CJK}
\usepackage{indentfirst}
\usepackage{multirow}
\usepackage{epstopdf}
\usepackage{graphicx}
\usepackage{footmisc}
\usepackage{amsfonts}
\usepackage{mathrsfs}
\usepackage{setspace}
\usepackage{algorithm,amsbsy,amsmath,amssymb,epsfig,bbm,mathrsfs, bbm} 
\usepackage{amsthm}
\usepackage{verbatim}
 \usepackage{geometry}
\usepackage[subfigure]{graphfig}

\newtheorem{theorem}{Theorem}

\allowdisplaybreaks

\begin{document}

\title{\vspace{-5mm} A Bandit Approach for Mode Selection in Ambient Backscatter-Assisted Wireless-Powered Relaying}

\author{Guangxia Li, Xiao Lu, Dusit Niyato

 \vspace{-5mm}
\thanks{G. Li is with the School of Computer Science and Technology, Xidian University, China, X. Lu is with the Department of Electrical $\&$ Computer Engineering, University of Alberta, Canada, D. Niyato is with the School of Computer Science and Engineering, Nanyang Technological University, Singapore.}  
}

{}

\maketitle

\begin{abstract}
Backscattering assisted wireless-powered communication combines ultralow-power backscatter transmitters with energy harvesting devices.
This paper investigates the transmission mode selection problem of a hybrid relay that forwards data by switching between the active wireless-powered transmission and the passive ambient backscattering.
It first presents a hybrid relay system model and derives its end-to-end success probability under theoretically optimal, but practically unrealistic, conditions.
The transmission mode selection is then formulated as a stochastic two-armed bandit problem in a varying environment where the distributions of rewards are nonstationary.
The proposed model selection scheme does not assume to have access to any channel states or network conditions, but merely relies on learning from past transmission records.
Numerical analyses are performed to validate the proposed bandit-based mode selection approach.
\end{abstract}

\begin{IEEEkeywords}
wireless-powered relaying, ambient backscatter, mode selection, bandit
\end{IEEEkeywords}

\IEEEpeerreviewmaketitle

\section{Introduction}
\label{sec1}

Wireless-powered communication can replenish the energy storage of devices remotely by means of radio-frequency (RF) energy harvesting techniques \cite{lu2014wireless,lu2015wireless}.
Due to its active transmission nature, a wireless-powered transmitter consumes relatively high circuit power to generate RF signals, and thus demands sufficient energy supply from the environment.
A promising solution is to integrate wireless-powered communication with a passive communication function to build a hybrid system so that the system's sustainability can be extended, even when the harvested energy is scarce~\cite{lu2017analysis}.
As a practical passive communication paradigm, ambient backscattering transmits data by reflecting nearby TV, WiFi and cellular RF signals.
It consumes ultralow energy since it does not generate RF signals actively.
This makes ambient backscattering an ideal complement to wireless-powered communication in real-world applications~\cite{DBLP:journals/wc/LuN0KXH18}.

A critical issue for a hybrid transmitter is the switch between two transmission modes, i.e., the wireless-powered communication and the ambient backscattering, under varying network conditions.
The so-called transmission mode selection problem has been extensively studied in multichannel communication systems, mostly for the cellular network with underlying device-to-device communications.
It is mainly solved by optimization techniques that are inclined to be computationally intensive or heuristic methods where a large amount of information (e.g., periodic collections of channel state information (CSI)) is required for decision making~\cite{DBLP:journals/comsur/AsadiWM14,niyato2015game,niyato2016distributed}.
For backscatter-assisted communication, however, the mode selection problem has not yet been thoroughly examined.
A representative work selects ambient backscattering and wireless-powered transmission based on the power and signal-to-noise ratio (SNR) thresholds, which are dependent upon the CSI and environmental factors, such as the distribution and the transmission load of ambient energy sources~\cite{DBLP:journals/twc/LuJNKH18}.

Since the instantaneous CSI used for model selection always incurs a communication overhead and channel state estimation errors are unavoidable in practice, there is a demand for lightweight approaches that are applicable to systems with energy constraints and limited computing power.
We thus consider the mode selection in a hybrid relay system as making decisions with uncertainty and formulate it as a stochastic two-armed bandit game.
A two-armed bandit problem can be analogous to a lever-operated slot gambling machine where a player can choose to pull either the left or the right arm (in the case of mode selection, wireless-powered communication or ambient backscattering), each giving a random reward with the distribution unknown to the player.
The goal is to maximize the total reward, which depends on the actions.

In this paper, we focus on study the mode selection problem for a hybrid relaying system. The main technical contributions are summarized as follows. 
\begin{itemize}
\item 
We first derive analytical expressions to characterize the end-to-end success probability of the relay under optimal conditions, which can serve as an upper bound of the system's coverage performance.

\item We then adopt a state-of-the-art bandit policy --- KL-UCB~\cite{DBLP:journals/jmlr/GarivierC11} to devise a practical mode selection method that requires no prior information about channel states or network conditions, but solely relies on the past transmission records.

\item  
We further tailor the policy with a discount factor that promotes recent records to make the model more robust to the changing environment.

\end{itemize}

To our knowledge, this work is the first attempt to solve the mode selection problem using the bandit approach for the integrated wireless powered communications with ambient backscattering.

\section{System Model and Mode Selection Scheme}
\label{sec2}

\begin{figure}
\centering
\includegraphics[width=0.45 \textwidth]{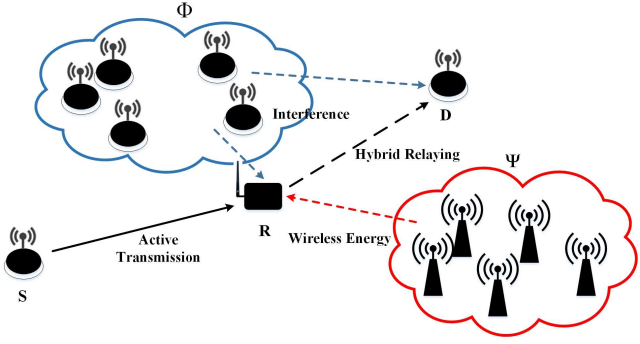} 
\caption{Hybrid relaying system model.} \vspace{-8mm}
\label{fig1}
\end{figure}

As shown in Fig.~\ref{fig1}, we consider a two-hop transmission system consisting of a source node $\mathrm{S}$, a destination node $\mathrm{D}$, and an energy-harvesting-equipped relay node $\mathrm{R}$ for forwarding data from $\mathrm{S}$ to $\mathrm{D}$, similar to the system models in \cite{lu2018performance,lu2019ambient}.
The relay $\mathrm{R}$ can switch between wireless-powered communication (referred to as the active relay mode, and denoted by $\mathrm{A}$) and ambient backscattering (the passive relay mode, $\mathrm{P}$), and consequentially consumes different circuit power.
Let $E_\mathrm{A}$ and $E_\mathrm{P}$ denote the circuit power consumption in a time slot when $\mathrm{R}$ works in the active mode and passive mode, respectively.
It is assumed that $E_\mathrm{A} >> E_\mathrm{P}$ since the active mode always consumes much more power than the passive mode in practice.

Around the hybrid relay $\mathrm{R}$ are a bunch of ambient transmitters that fall into two categories.
The transmitters denoted as $\Psi$ in Fig.~\ref{fig1} are considered to have high transmit power, and thus are more suitable to be the signal source for energy harvesting and backscattering.
The transmitters in $\Phi$ are counterparts of the source node $\mathrm{S}$.
If the hybrid relay $\mathrm{R}$ works in the active mode, the $\mathrm{R}$-to-$\mathrm{D}$ transmission will be carried out on the same frequency band of $\mathrm{S}$ and will be impaired by the interference from $\Phi$.
Otherwise, if the relay works in the passive mode, it will perform ambient backscattering by using the signal from $\Phi$.
To account for the spatial randomness of ambient transmitters $\Psi$ and $\Phi$, we assume that their distributions follow independent homogeneous Poisson point processes (PPPs).
In addition, we denote the transmit power of $\Psi$ ($\Phi$) as $\widetilde{P}_T$ ($P_T$) and the spatial density of $\Psi$ ($\Phi$) as $\widetilde{\zeta}$ ($\zeta$).

The system operates on a time-slot basis.
For each time-slot, the first $\eta$ fraction is allocated for the relay $\mathrm{R}$ to harvest energy.
Then, the first and the second halves of the remaining time-slot are allocated for the $\mathrm{S}$-to-$\mathrm{R}$ transmission and the $\mathrm{R}$-to-$\mathrm{D}$ transmission, respectively.
The power of the incident RF signals at $\mathrm{R}$ from $\Psi$ in an entire time-slot can be calculated as $Q_{\mathrm{R}} = \widetilde{P}_T \sum_{\i \in \Psi} | h_{i,\mathrm{R}} |^2 d_{i,\mathrm{R}}^{-\widetilde{\alpha}}$, where $h_{a,b}$ and $d_{a,b}$ represent the amplitude gain of the channel and the distance between $a$ and $b$, respectively, and $\widetilde{\alpha}$ represents the path-loss exponent for the signals from $\Psi$.
The amount of harvested energy in a time-slot can thus be represented as $E_{\mathrm{R}} = \eta \beta Q_{\mathrm{R}}$, where $\beta$ denotes the RF-to-DC energy conversion efficiency of relay $\mathrm{R}$.

Denote the transmit power of source node $\mathrm{S}$ as $P_\mathrm{S}$.
During the $\mathrm{S}$-to-$\mathrm{R}$ transmission phase, the received signal-to-interference-plus-noise ratio (SINR) at relay $\mathrm{R}$ is
\begin{equation}
\nu^{\mathrm{R}} = \frac{ P_{\mathrm{S}} | h_{\mathrm{S},\mathrm{R}} |^2 d^{-\alpha}_{\mathrm{S},\mathrm{R}} }{ I_{\mathrm{R}} +\sigma^{2} },
\label{eq:SINR_R}
\nonumber
\end{equation}
where $\alpha$ is the path-loss exponent for the signals from $\Phi$, $I_{\mathrm{R}}=\sum_{i \in \Phi} P_{T} | h_{i,\mathrm{R}} |^2 d^{-\alpha}_{i,\mathrm{R}}$ is the aggregated interference power at $\mathrm{R}$,  and $\sigma^2$ is the power of the additive white Gaussian noise (AWGN) in the transmission frequency of $\Phi$.

Since the energy harvested by the relay $\mathrm{R}$ is limited, they are first used to power the circuit in the active mode. 
The surplus energy (if any) is then reserved in energy storage and used to perform the wireless-powered transmission during the $\mathrm{R}$-to-$\mathrm{D}$ transmission phase later on.
Let $E_{C}$ denote the capacity of the energy storage of relay $\mathrm{R}$.       
The transmit power of relay $\mathrm{R}$ during the $\mathrm{R}$-to-$\mathrm{D}$ transmission phase is
\begin{equation}
P_{\mathrm{A}} =
\begin{cases}
\frac{2 E_{C}}{1-\eta} & \text{if } E_{\mathrm{R}} \geq  E_{\mathrm{A}} + E_{C}\\
\frac{ 2 (\eta \beta Q_{\mathrm{R}} - E_{\mathrm{A}}) }{ 1 - \eta } & \text{if } E_{\mathrm{A}}+ E_{C}  > E_{\mathrm{R}} \geq E_{\mathrm{A}}\\
0 & \text{otherwise}.
\end{cases}
\label{eq:transmit_power_RA}
\end{equation}  

The received SINR at the destination node $\mathrm{D}$ in the active mode can thus be calculated as
\begin{equation}
\nu^{\mathrm{D}}_{\mathrm{A}} = 
\frac{ P_{\mathrm{A}} | h_{\mathrm{R}, \mathrm{D}} |^2 d^{-\alpha}_{\mathrm{R}, \mathrm{D}} }
     { I_{\mathrm{D}} + \sigma^2 }.
\label{eq5}
\nonumber
\end{equation} 
where $I_{\mathrm{D}}=\sum_{i \in \Phi } P_T | h_{i, \mathrm{D}} |^2 d^{-\alpha}_{i, \mathrm{D}}$ is the aggregated interference power at $\mathrm{D}$ in the active mode.

In the passive mode, the backscattered power of relay $\mathrm{R}$ is $P_{\mathrm{P}} = \Gamma \xi Q_R$, where $\Gamma$ denotes the fraction of the RF signals reflected during backscattering, and $\xi$ is the backscatter efficiency representing the portion of the reflected signals that are effectively used to carry the modulated data.
In general, the signals from the carrier emitters will cause self-interference for the backscattered signals.
However, the relay $\mathrm{R}$ is considered to employ a physical-layer technique, namely, frequency shifting~\cite{zhang2016enabling}, to eliminate the self-interference from the carrier emitters.
It is then assumed that the relay $\mathrm{R}$ can remove the self-interference completely.
The received SNR at the destination node $\mathrm{D}$ can be calculated as
\begin{equation}
\nu^{\mathrm{D}}_{\mathrm{P}} = \frac{ P_{\mathrm{P}} | h_{\mathrm{R},\mathrm{D}} |^2 d^{-\alpha}_{\mathrm{R},\mathrm{D}} }{ \widetilde{\sigma}^2 },
\label{eq7}
\nonumber
\end{equation}
where $\widetilde{\sigma}^2$ represents the power of the AWGN in the transmission frequency of $\Psi$.
\section{Success Probability with Optimal Mode Selection Policy}
\label{sec3}

The aforementioned hybrid relaying system is considered to make a successful transmission if (1) the relay $\mathrm{R}$ can harvest enough energy to support its circuit operation; (2) during the $\mathrm{S}$-to-$\mathrm{R}$ transmission phase, the achieved SINR at the relay, i.e., $\nu^{\mathrm{R}}$, is greater than a threshold $\tau_{\mathrm{A}}$; and (3) during the $\mathrm{R}$-to-$\mathrm{D}$ transmission phase, depending on the selected relay mode (active or passive), the achieved SINR $\tau^{\mathrm{D}}_{\mathrm{A}}$ or SNR $\tau^{\mathrm{D}}_{\mathrm{P}}$ at the destination node is greater than a threshold $\tau_{\mathrm{A}}$ or $\tau_{\mathrm{P}}$, respectively.
Let $\mathrm{M} \in \{\mathrm{A}, \mathrm{P}\}$ be an indicator for the selected mode.
The success probability of the hybrid relaying system in the mode $\mathrm{M}$ can be represented as
\begin{equation}
\mathcal{P}_{\mathrm{M}} = \mathbb{P}[ E_{\mathrm{R}} > E_{\mathrm{M}}, \nu^{\mathrm{R}} > \tau_{\mathrm{A}}, \nu^{\mathrm{D}}_{\mathrm{M}} > \tau_{\mathrm{M}} ].
\label{def:CP}
\nonumber
\end{equation}

It is obvious that if the hybrid relay is fully aware of the channel and network conditions, it can infer the transmission qualities of the two modes easily.
If any mode or both modes can achieve a successful transmission, the relay can select that particular mode or either mode, respectively.
Even through this setting is not realistic in practice (since obtaining CSI is costly, if not impossible), the coverage probability derived under it is still useful since it serves as the theoretical upper bound of the hybrid relay's performance.
We thus characterize the coverage probability with the optimal mode selection using a theorem.
\begin{theorem}
The success probability of the hybrid relaying with the optimal mode selection is given by
\begin{align}
\mathcal{P}^{opt}_{\textup{HR}} = \;
& \mathcal{P}_{\mathrm{A}} + \mathcal{J} + 
\exp(-g_{1} \sigma^2) \mathcal{L}_{I_{\mathrm{R}} } \big( g_{1} \big) \nonumber \\
& \times \int^{B_{3}}_{ B_{2}}  \!\!\!
\exp \! \big ( \!-\!  g_{3}(q) \widetilde{\sigma}^2 \big ) 
f_{Q_{\mathrm{R}}}(q) \mathrm{d} q,
\label{HR_opt}
\end{align}
where $\mathcal{P}_{\mathrm{A}}$ and $\mathcal{J}$ are given by Eq.~\eqref{PA} and Eq.~\eqref{MJ}, respectively,
$B_{1}    = \frac{ E_{C} }{ \eta \beta }$,
$B_{2}    = \frac{ E_{\mathrm{P}} }{ \eta \beta }$,
$B_{3}    = \frac{ E_{\mathrm{A}} }{ \eta \beta }$,
$g_{1}    = \frac{ d^{\alpha}_{\mathrm{S}, \mathrm{R}} \tau_{\mathrm{A}} }{ P_{\mathrm{S}} }$, 
$g_{2}(q) = \frac{ d^{\alpha}_{\mathrm{R}, \mathrm{D}} \tau_{\mathrm{A}} (1 - \eta) }{ 2(q - E_{\mathrm{A}}) }$,
$g_{3}(q) = \frac{ d^{\alpha}_{\mathrm{R}, \mathrm{D}} \tau_{\mathrm{P}} }{ \Gamma \xi q }$,
$f_{Q_{\mathrm{R}}}(q) = \mathcal{L}^{-1} \big \{ L(s, \widetilde{P}_T, \widetilde{\zeta}) \big \} (q)$, 
\begin{figure*}
\begin{align}
& \mathcal{P}_{\mathrm{A}} \! = \exp \! \big( \! - \! g_{1} \sigma^2 \big) \! \! 
\int^{B_{1}}_{B_{3}} \!\!\! \exp \big( \! - g_{2} \big( \eta \beta q \big) \sigma^2 \big) 
\mathcal{L}_{I_{\mathrm{R}},I_{\mathrm{D}}} \big( g_{1}, g_{2} \big( \eta \beta q \big) \big) 
f_{Q_{\mathrm{R}}} (q) \mathrm{d} q  \nonumber\\
& \hspace{80mm} + 
\exp \! \big( \! - \! g_{2} \big(E_{C}\big) \sigma^2 \big) 
\mathcal{L}_{I_{\mathrm{R}},I_{\mathrm{D}}} \big( g_{1}, g_{2} \big(E_{C}\big) \! \big) 
\big( 1- \bar{F}_{Q_{\mathrm{R}}} (B_{1}) \big), 
\label{PA}
\vspace{-6mm}
\end{align} \hrulefill
\end{figure*}
\begin{figure*} \vspace{-5mm}
\begin{align}
\mathcal{J} \! = 
 &\exp  \! \big(\!- g_{1} 
 \sigma^2 \big) \! \bigg( \!   \int^{ B_{1} }_{B_{3}}  \! \!\! \exp \! \big( \! - g_{2} (\tau_{\mathrm{A}}, \eta \beta q ) \sigma^2 \big) \!   
 \Big(    \mathcal{L}_{I_{\mathrm{R}}} \big( g_{1} 
 \big) \!-\! \mathcal{L}_{I_{\mathrm{R}},I_{\mathrm{D}} } \big( g_{1}
 , g_{2} (\tau_{\mathrm{A}}, \eta \beta q )  \big)  \exp(- \sigma^2 g_{2} (\tau_{\mathrm{A}}, \eta \beta q )  )   \!  \Big) f_{Q_{\mathrm{R}}} (q) \mathrm{d} q   \nonumber \\ 
 & \hspace{4mm}    +         \Big(    \mathcal{L}_{I_{\mathrm{R}}} \big( g_{1} 
 \big)  
  - \mathcal{L}_{I_{\mathrm{R}},I_{\mathrm{D}} } \big( g_{1} 
  , g_{2} (\tau_{\mathrm{A}}, E_{C} ) \big) \exp(- \sigma^2 g_{2} (\tau_{\mathrm{A}}, E_{C} )   )     \Big)   
  \big( 1- F_{Q_{\mathrm{R}}}   ( B_{1} ) \big) \exp \big( \! - g_{2} (\tau_{\mathrm{A}}, E_{C} ) \sigma^2 \big)   \bigg), 
\label{MJ}
\vspace{-6mm}
\end{align} \vspace{-5mm} \hrulefill 
\end{figure*}
$L$ is given by
\begin{align}
L(s, p, z) = \exp \Big( \!\! - \! \frac{2}{\alpha} \pi^2 z(s p)^{\frac{2}{\alpha}} \mathrm{csc} \big( \frac{2 \pi}{\alpha} \big) \Big),
\label{LTIR}
\end{align}
and $\mathcal{L}_{I_{\mathrm{R}},I_{\mathrm{D}}} (s_{1},s_{2})$ is given by
\begin{align}
& \mathcal{L}_{I_{\mathrm{R}},I_{\mathrm{D}}} (s_{1},s_{2}) = 
\exp \Big( \!  - \zeta \! \int^{\infty}_{0} \!\!\! \int^{2\pi}_{0} \! \Big[ 1 - (1 + s_{1} P_{T}  r^{-\alpha} )^{-1} \nonumber \\
& \hspace{25mm} \times \big( 1 + \! s_{2} P_{T} l(r,\theta )^{-\alpha} \big)^{\! -1} \Big] r \mathrm{d} \theta \mathrm{d} r \Big),
\label{CP_DF}
\end{align}
where $l(r, \theta) = \big(r^2+d^2_{\mathrm{R},\mathrm{D}} - 2 r d_{\mathrm{R},\mathrm{D}} \cos(\theta) \big)^{\frac{1}{2}}$.
\label{theorem-opt}
\end{theorem}

\begin{figure*}[htp]
\begin{align}
& \mathcal{P}^{opt}_{\textup{HR}}  = \mathbb{P} \big[ \{ S_{\mathrm{A}} \}  \cup  \{ S_{\mathrm{P}} \} \big] = \mathbb{P} \big[  S_{\mathrm{A}}  \big] + \mathbb{P} \big[  S_{\mathrm{P}}  \big] - \mathbb{P} \big[ \{ S_{\mathrm{A}} \}  \cap  \{ S_{\mathrm{P}} \} \big] \nonumber \\
& = \mathbb{P} \big[ \nu^{\mathrm{R}}  > \tau_{\mathrm{A}}, \nu^{\mathrm{D}}_{\mathrm{A}} > \tau_{\mathrm{A}}, E_{\mathrm{R}} > E_{\mathrm{A}} \big]  + \mathbb{P} \big[ \nu^{\mathrm{R}}  > \tau_{\mathrm{A}}, \nu^{\mathrm{D}}_{\mathrm{P}} > \tau_{\mathrm{P}}, E_{\mathrm{R}} > E_{\mathrm{P}} \big]  - \mathbb{P} \big[ \nu^{\mathrm{R}}  > \tau_{\mathrm{A}}, \nu^{\mathrm{D}}_{\mathrm{A}} > \tau_{\mathrm{A}}, \nu^{\mathrm{D}}_{\mathrm{P}} > \tau_{\mathrm{P}}, E_{\mathrm{R}} > E_{\mathrm{A}} \big]  \nonumber \\
& = \underbrace{ \mathbb{P} \big[ \nu^{\mathrm{R}}  > \tau_{\mathrm{A}}, \nu^{\mathrm{D}}_{\mathrm{A}} > \tau_{\mathrm{A}}, E_{\mathrm{R}} > E_{\mathrm{A}} \big] }_{:=\mathcal{P}_{\mathrm{A}} }  + \underbrace{\mathbb{P} \big[ \nu^{\mathrm{R}}  > \tau_{\mathrm{A}}, \nu^{\mathrm{D}}_{\mathrm{P}} > \tau_{\mathrm{P}}, \nu^{\mathrm{D}}_{\mathrm{A}} \leq \tau_{\mathrm{A}}, E_{\mathrm{R}} > E_{\mathrm{A}} \big]}_{:=\mathcal{J}}  + \underbrace{\mathbb{P} \big[ \nu^{\mathrm{R}}  > \tau_{\mathrm{A}},   \nu^{\mathrm{D}}_{\mathrm{P}} > \tau_{\mathrm{P}}, E_{\mathrm{A}} >  E_{\mathrm{R}} > E_{\mathrm{P}} \big]}_{:=\mathcal{K}}.  
\label{eqn:HR_OPT} \vspace{-5mm}
\end{align}  
\hrulefill 
\end{figure*}
\begin{proof} 
Let $S_{\mathrm{A}}$ and $S_{\mathrm{P}}$ denote the events that the hybrid relay achieves successful transmission in the active mode and the passive mode, respectively.
According to the selection criteria introduced above, the success probability of hybrid relaying with the optimal mode selection can be expressed as Eq.~\eqref{eqn:HR_OPT} shown on the top of the next page.
We first derive $\mathcal{P}_{\mathrm{A}}$ as
\begin{align}
 &\hspace{0mm} \mathcal{P}_{\mathrm{A}}4    = \mathbb P [ \nu^{\mathrm{R}} > \tau_{\mathrm{A}}, \nu^{\mathrm{D}}_{\mathrm{A}} \! > \! \tau_{\mathrm{A}}, E_{\mathrm{R}}  >  E_{\mathrm{A}}   ]\nonumber \\ 
&\hspace{0mm} = \mathbb{P} \bigg[ h_{\mathrm{S},\mathrm{R}}   \!\geq\!  \frac{ \tau_{\mathrm{A}}   (I_{\mathrm{R}}\!+\!\sigma^2)   }{d^{-\alpha}_{\mathrm{S},\mathrm{R}}  P_{\mathrm{S}} },  h_{\mathrm{R},\mathrm{D}} \!\geq\! \frac{ \tau_{\mathrm{A}}   (I_{\mathrm{D}}\!+\!\sigma^2)   }{  d^{-\alpha}_{\mathrm{R},\mathrm{D}} P_{\mathrm{A}}  
 }, Q_{\mathrm{R}} \! \geq \! \frac{ E_{\mathrm{A} }    }{ \eta \beta  }   \bigg] \nonumber \\
 &\hspace{0mm} \overset{(a)}{=} \! \mathbb{E}_{ I_{\mathrm{R}},I_{\mathrm{D}} } \! \Bigg[\!\exp \! \bigg( \!\!-\! \frac{ \tau_{\mathrm{A}}   (I_{\mathrm{R}}\!+\!\sigma^2)   }{ d^{-\alpha}_{\mathrm{S},\mathrm{R}} P_{\mathrm{S}} } \!  \bigg)    \nonumber \\
  & \hspace{30mm} \times \exp \! \bigg( \!-\! \frac{  \tau_{\mathrm{A}}   (I_{\mathrm{D}}\!+\!\sigma^2)   }{  d^{-\alpha}_{\mathrm{R},\mathrm{D}} P_{\mathrm{A}}  
    }\! \bigg) \mathbbm1_{ \! \big\{ Q_{\mathrm{R}}   \geq   \frac{ E_{\mathrm{A} }    }{ \eta \beta  } \big \}}  
 \!\Bigg]   \nonumber\\ 
 &\hspace{0mm} \overset{(b)}{=} \exp \! \bigg(  \!\!-\!    \frac{d^{\alpha}_{\mathrm{S},\mathrm{R}}\tau_{\mathrm{A}}  \sigma^2    }{  P_{\mathrm{S}} } \!\bigg) \! \int^{\infty}_{0} \! \exp \bigg( \!    - \frac{ d^{\alpha}_{\mathrm{R},\mathrm{D}} \tau_{\mathrm{A}}   \sigma^2    }{  P_{\mathrm{A}}  
    } \bigg) \nonumber \\
 &  \hspace{30mm} \times \mathcal{L}_{I_{\mathrm{R}},I_{\mathrm{D}}}  \bigg(  \! \frac{d^{\alpha}_{\mathrm{S},\mathrm{R}}\tau_{\mathrm{A}}    }{  P_{\mathrm{S}} } , \frac{ d^{\alpha}_{\mathrm{R},\mathrm{D}} \tau_{\mathrm{A}}  }{  P_{\mathrm{A}}  
      } \! \bigg) f_{Q_{\mathrm{R}}} (q) \mathrm{d} q   \nonumber \\ 
    & \overset{(c)}{=}  \exp \! \bigg(  \!\!-\!    \frac{d^{\alpha}_{\mathrm{S},\mathrm{R}}\tau_{\mathrm{A}}  \sigma^2    }{  P_{\mathrm{S}} } \!\bigg) \! \int^{\frac{E_{C}  }{\beta \eta } }_{\frac{E_{A}  }{\beta \eta }} \! \exp \bigg( \!    - \frac{ d^{ \alpha}_{\mathrm{R}, \mathrm{D} }  \tau_{\mathrm{A}} (  1 -  \eta ) \sigma^2 }{ 2  ( \eta \beta q -E_{\mathrm{A}}) }       \bigg) \nonumber \\
    &  \hspace{15mm} \times \mathcal{L}_{I_{\mathrm{R}},I_{\mathrm{D}}}  \bigg(  \! \frac{d^{\alpha}_{\mathrm{S},\mathrm{R}}\tau_{\mathrm{A}}    }{  P_{\mathrm{S}} } ,  \frac{ d^{ \alpha}_{\mathrm{R}, \mathrm{D} }  \tau_{\mathrm{A}} (  1 -  \eta ) \sigma^2 }{ 2  ( \eta \beta q -E_{\mathrm{A}}) }  \! \bigg) f_{Q_{\mathrm{R}}} (q) \mathrm{d} q   \nonumber \\ 
    & + \exp \! \Big( \! -  \!   \frac{ d^{ \alpha}_{\mathrm{R}, \mathrm{D} }  \tau_{\mathrm{A}} (  1 -  \eta ) \sigma^2 }{ 2  ( E_{C} -E_{\mathrm{A}}) }    \Big) 
          \mathcal{L}_{I_{\mathrm{R}},I_{\mathrm{D}}} \! \Big( \frac{d^{\alpha}_{\mathrm{S},\mathrm{R}}\tau_{\mathrm{A}}    }{  P_{\mathrm{S}} }  ,  \frac{ d^{ \alpha}_{\mathrm{R}, \mathrm{D} }  \tau_{\mathrm{A}} (  1 -  \eta )   }{ 2  ( E_{C} -E_{\mathrm{A}}) }   \!  \Big)  \nonumber \\
          & \hspace{50mm} \times \big( 1- F_{Q_{\mathrm{R}}} (B_{1}) \big), 
  \label{eqn:CP_DF}  
\end{align}
where $(a)$ follows because $h_{\mathrm{S},\mathrm{R}}$ and $h_{\mathrm{R},\mathrm{D}}$  are independent and exponentially distributed, $(b)$
defines the Laplace transform of the joint PDF of $I_{\mathrm{R}}$ and $I_{\mathrm{D}}$, i.e., $\mathcal{L}_{I_{\mathrm{R}},I_{\mathrm{D}}} (s_{1},s_{2}) : = \mathbb{E} \big[ \exp( - s_{1} I_{\mathrm{R}} - s_{2}  I_{\mathrm{D}} ) \big] $, and $(c)$ inserts the expression of $P_{\mathrm{A}}$ in (\ref{eq:transmit_power_RA}). Moreover, after applying the substitutions $g_{1}= \frac{d^{ \alpha}_{\mathrm{S}, \mathrm{R} }\tau_{\mathrm{A}}}{ P_{\mathrm{S}}  } $ and  $ g_{2} (v,p) = \frac{ d^{ \alpha}_{\mathrm{R}, \mathrm{D} }  v (  1 -  \eta )  }{ 2  (p-E_{\mathrm{A}}) }$, we have  $\mathcal{P}_{\mathrm{A}}$ expressed as (\ref{PA}), where  $f_{Q_{\mathrm{R}}}(q)$ and $F_{Q_{\mathrm{R}}}(q)$ are the PDF and CDF of $Q_{\mathrm{R}}$, respectively.  
$f_{Q_{\mathrm{R}}}(q)$ can be obtained by taking the inverse Laplace transform of the Laplace transform of $Q_{\mathrm{R}}$ as follows   
\begin{align}
f_{Q_{\mathrm{R}}}(q) 
& = \mathcal{L}^{-1} \bigg \{ \mathbb{E} \Big[  \exp \Big( \! - s \sum_{i \in \Psi} \widetilde{P}_{T} | h_{i,\mathrm{R}} |^2 d^{-\widetilde{\alpha}}_{i,\mathrm{R}} \Big) \Big]  \bigg \} (q) \nonumber \\
& \overset{(d)}{=} \mathcal{L}^{-1} \bigg \{ \exp \! \bigg( \!\! - \! 2 \pi \widetilde{\zeta} \! \int^{\infty}_{0} \!\! \Big (   1 \! -  \Big( 1 \!+ \! \frac{ s \widetilde{P}_{\mathrm{T}}      }{   r^{ \widetilde{\alpha}}     }      \Big)^{ \!\!-1}  \Big ) r \mathrm{d} r   \! \bigg) \bigg\} (q) \nonumber \\
&  =  \mathcal{L}^{-1} \Big \{ \exp \Big( \!\! - \! \frac{2}{\widetilde{\alpha}} \pi^2 \widetilde{\zeta}  ( s \widetilde{P}_{T} )^{\frac{2}{\widetilde{\alpha}}} \mathrm{csc} \big( \frac{2 \pi}{\widetilde{\alpha}} \big)  \Big) \Big \} (q),
\label{QR_pdf}
\end{align}
where $(d)$ applies the probability generating functional (PGFL) for a PPP.
Then, by integrating the PDF in Eq.~\eqref{QR_pdf}, we have the CDF of $Q_{\mathrm{R}}$ as $F_{Q_{\mathrm{R}}}(q)=\mathcal{L}^{-1} \big \{ \frac{1}{s} \exp \big( \! - \! \frac{2 \pi^2}{\alpha}  \widetilde{\zeta} ( s P_{T} )^{\frac{2}{\widetilde{\alpha}}} \mathrm{csc} \big( \frac{2 \pi}{\alpha} \big)  \big) \big \} (q)$.
Moreover, $\mathcal{L}_{ I_{\mathrm{R}},I_{\mathrm{D}} } (s_{1},s_{2})$  can be derived as
\begin{align}
 &  \hspace{-3mm} \mathcal{ L}_{ I_{\mathrm{R}},I_{\mathrm{D}} } (s_{1},s_{2})  =  \mathbb{E} \big[ \exp( - s_{1} I_{\mathrm{R}} - s_{2}  I_{\mathrm{D}} ) \big] \nonumber \\
 &   \hspace{-3mm} 
 = \mathbb{E}_{\Phi} \bigg[    \prod_{i \in \Phi }  \bigg(   1 +    \!   s_{1} P_{T}      d^{-\alpha}_{i,\mathrm{R}}   \bigg)^{\!\!-1} \Big(   1 +      \! \ s_{2} P_{T}     d^{-\alpha}_{i,\mathrm{D}}   \Big)^{\!\!-1} \bigg] \nonumber \\
 & \hspace{-3mm} \overset{(e)}{=}   \exp \! \bigg( \!\! - \!  \zeta \! \int^{\infty}_{0} \!\!\! \int^{2\pi}_{0} \! \bigg[   1 \! - \Big(   1 +  \!  s_{1}   P_{T}      r^{-\alpha}   \Big)^{\!\!-1}  \nonumber \\
 & \hspace{-3mm} \times  \bigg(\! 1 + \!  s_{2}  P_{T}     \big(r^2+d^2_{\mathrm{R},\mathrm{D}} - 2 r d_{\mathrm{R},\mathrm{D}} \cos(\theta) \big)^{-\frac{\alpha}{2}}    \! \bigg)^{\!\!\!-1}  \bigg] r \mathrm{d} \theta \mathrm{d} r   \! \bigg), \label{jointLP} 
\end{align}
where $(e)$ takes the average over the independent exponential random variables $h_{i,\mathrm{R}}$ and $h_{i,\mathrm{D}}$.

Next, we derive $\mathcal{J}$ as in (\ref{J}),
\begin{align}
   &  \mathcal{J}  =  \mathbb{P} \big[ \nu^{\mathrm{R}} > \tau_{\mathrm{A}}, \nu^{\mathrm{D}}_{\mathrm{P}} > \tau_{\mathrm{P}}, \nu^{\mathrm{D}}_{\mathrm{A}} \leq \tau_{\mathrm{A}}, E_{\mathrm{R}} > E_{\mathrm{A}} \big] \nonumber \\ &  
   \overset{(f)}{=}  \mathbb{E} \Bigg[\!
\exp \! \bigg(\!-\! \frac{d^{\alpha}_{\mathrm{S},\mathrm{R}} \tau_{\mathrm{A}} (I_{\mathrm{R}}\!+\!\sigma^2)}{P_{\mathrm{S}}} \! \bigg) 
\exp \! \bigg(\!-\! \frac{d^{\alpha}_{\mathrm{R},\mathrm{D}} \tau_{\mathrm{P}} \widetilde{\sigma}^2}{\Gamma \xi Q_{\mathrm{R}}} \bigg) \nonumber \\
& \hspace{5mm} \times \bigg[ 1- \exp \! \bigg(\!-\! \frac{d^{\alpha}_{\mathrm{R},\mathrm{D}} \tau_{\mathrm{A}} (I_{\mathrm{D}} \!+\! \sigma^2)}{P_{\mathrm{A}}} \! \bigg) \bigg] 
\mathbbm1_{ \{ E_{\mathrm{R}} > E_{\mathrm{A}} \} }  
\!\Bigg], 
\label{J} 
\end{align}
where $(f)$ follows because $h_{\mathrm{S},\mathrm{R}}$ and $h_{\mathrm{R},\mathrm{D}}$ are independent and exponentially distributed. Then, by inserting the expression of $P_{\mathrm{A}}$ in (\ref{eq:transmit_power_RA})  and applying the substitutions    $g_{1}= \frac{d^{ \alpha}_{\mathrm{S}, \mathrm{R} }\tau{\mathrm{A}}}{ P_{\mathrm{S}}  } $ and  $ g_{2} (v,p) = \frac{ d^{ \alpha}_{\mathrm{R}, \mathrm{D} }  v (  1 -  \eta )  }{ 2  (p-E_{\mathrm{A}}) }$,  we have $\mathcal{J}$ expressed as (\ref{MJ}).


Similarly, we have $\mathcal{K}$ derived as in (\ref{K}), 
\begin{figure*}
\vspace{-3mm}
\begin{align} \label{K}
& \mathcal{K}  = \mathbb P [ \nu^{\mathrm{R}} > \tau_{\mathrm{A}}, \nu^{\mathrm{D}}_{\mathrm{P}} \! > \! \tau_{\mathrm{P}}, E_{\mathrm{R}}  >  E_{\mathrm{P}}   ] 
= \mathbb{P} \bigg[ h_{\mathrm{S},\mathrm{R}}   \!\geq\!  \frac{d^{\alpha}_{\mathrm{S},\mathrm{R}}\tau_{\mathrm{A}}   (I_{\mathrm{R}}\!+\!\sigma^2)   }{  P_{\mathrm{S}} },  \widetilde{h}_{\mathrm{R},\mathrm{D}} \!\geq\! \frac{ d^{\alpha}_{\mathrm{R},\mathrm{D}} \tau_{\mathrm{P}}   \widetilde{\sigma}^2    }{  \Gamma \xi Q_{\mathrm{R}}  
 }, \frac{ E_{\mathrm{A} }    }{ \eta \beta  } \geq Q_{\mathrm{R}} \! \geq \! \frac{ E_{\mathrm{P} }    }{ \eta \beta  }   \bigg] \nonumber \\
 &  = \mathbb{E}_{ I_{\mathrm{R}}  } \! \Bigg[\!\exp \! \bigg( \!\!-\! \frac{ \tau_{\mathrm{A}}   (I_{\mathrm{R}}\!+\!\sigma^2)   }{ d^{-\alpha}_{\mathrm{S},\mathrm{R}} P_{\mathrm{S}} } \!  \bigg)  \exp \! \bigg( \!-\! \frac{  \tau_{\mathrm{P}}   \widetilde{\sigma}^2    }{  d^{-\alpha}_{\mathrm{R},\mathrm{D}} \Gamma \xi Q_{\mathrm{R}}   
  }  \bigg)  \mathbbm1_{ \frac{ E_{\mathrm{A} }    }{ \eta \beta  } \geq   \big\{ Q_{\mathrm{R}}   \geq   \frac{ E_{\mathrm{P} }    }{ \eta \beta  } \big \}}  
 \!\Bigg]   \nonumber\\
 &  =  \exp \! \bigg( \!\!-\! \frac{d^{\alpha}_{\mathrm{S},\mathrm{R}} \tau_{\mathrm{A}}   \sigma^2    }{  P_{\mathrm{S}} } \!  \bigg) \mathcal{L}_{I_{\mathrm{R}}} \bigg( \!   \frac{d^{\alpha}_{\mathrm{S},\mathrm{R}} \tau_{\mathrm{A}}     }{  P_{\mathrm{S}} } \!  \bigg)  \int^{\infty}_{ \frac{ E_{\mathrm{P} }    }{ \eta \beta  }}   \! \exp \! \bigg( \!-\! \frac{ d^{\alpha}_{\mathrm{R},\mathrm{D}} \tau_{\mathrm{P}}   \widetilde{\sigma}^2    }{   \Gamma \xi q    
   }  \bigg) f_{Q_{\mathrm{R}}}(q)    \mathrm{d} q ,
\end{align}  \vspace{-5mm} \hrulefill 
\end{figure*}
where 
$\mathcal{L}_{I_{\mathrm{R}}}(s) \! = \! \exp \Big( \!\! - \! \delta \pi^2 \zeta  \big(P_{T} s \big)^{\frac{2}{\alpha}} \mathrm{csc} \big( \pi \delta  \big) \! \Big) $ is the Laplace transform of $I_{\mathrm{R}}$ which can be obtained by following the derivation steps of (\ref{jointLP}).

Subsequently, according to (\ref{eqn:HR_OPT}), taking sum of $\mathcal{P}_{\mathrm{A}}$, $\mathcal{J}$ and $\mathcal{K}$ obtained in $(\ref{PA})$ $(\ref{MJ})$, $(\ref{K})$, respectively, and applying the substitutions
$B_{1} = \frac{E_{C}}{\eta\beta }$,
$B_{3} =\frac{E_{\mathrm{A}}}{\eta \beta}$, 
$g_{1} = \frac{d^{\alpha}_{\mathrm{S}, \mathrm{R}} \tau_{\mathrm{A}} }{P_{\mathrm{S}}}$,
$g_{2}(q) = \frac{d^{\alpha}_{\mathrm{R}, \mathrm{D}} \tau_{\mathrm{A}} (1 - \eta)}{2 (q - E_{\mathrm{A}})}$, and
$g_{3}(q) = \frac{d^{\alpha}_{\mathrm{R},\mathrm{D}} \tau_{\mathrm{P}}}{\Gamma \xi q}$ yield the final result in Theorem~\ref{theorem-opt}.
\end{proof}

\section{Mode Selection via Bandit}
\label{sec4}
\begin{table*}[t]
\small
\centering
\caption{Variable notations and values used by the simulation in Section~\ref{sec5}.}
\begin{tabular}{|l|l|l|} \hline
Symbol & Definition & Value in simulation\\ \hline
$d_{\mathrm{S},\mathrm{R}}$, $d_{\mathrm{R},\mathrm{D}}$ & Source-to-relay distance and relay-to-destination distance & $5\;\mathrm{m}$, $5\;\mathrm{m}$\\
$P_\mathrm{T}$, $\widetilde{P}_\mathrm{T}$ & Transmitter power of the interferers $\Phi$ and the carrier emitters $\Psi$ & $3\;\mathrm{dBm}$, $40\;\mathrm{dBm}$\\
$\zeta$, $\widetilde{\zeta}$ & Spatial density of the interferers $\Phi$ and the carrier emitters $\Psi$ & $0.001$, $0,001$\\
$P_\mathrm{S}$& Transmit power of the source node $\mathrm{S}$ during the $\mathrm{S}$-to-$\mathrm{R}$ transmission phase & $0.002$ Watt\\
 $P_\mathrm{A}$, $P_\mathrm{P}$ & Transmit power of the relay $\mathrm{R}$ in the active and the passive mode during &  \\
& the $\mathrm{R}$-to-$\mathrm{D}$ transmission phase & N. A.\\
$E_\mathrm{A}$, $E_\mathrm{P}$ & Circuit power consumption per time unit in the active and passive mode & $200\;\mathrm{\mu W}$, $10\;\mathrm{\mu W}$\\
$E_\mathrm{C}$ & Energy storage capacity of the relay & $0.002$ Joule\\
$\alpha$, $\widetilde{\alpha}$ & Pass-loss exponent for the interferers and the carrier emitters & $4$, $3$\\
$\beta$ & RF-to-DC conversion efficiency of the relay & $0.5$\\
$\Gamma$ & Backscatter coefficient parameter & $0.375$\\
$\eta$ & The portion of time in each time-slot used for energy harvesting & $0.4$\\
$\xi$ & Backscatter efficiency & $0.3$\\
$\sigma^2$, $\widetilde{\sigma}^2$ & Noise power for active transmission and passive transmission, respectively & $10^{-10}$, $10^{-9}$ Watt\\
$\tau_\mathrm{A}$ & Minimal required SINR at the destination node in the active mode & $0\;\mathrm{dB}$\\
$\tau_\mathrm{P}$ & Minimal required SNR at the destination node in the passive mode & $20\;\mathrm{dB}$\\
\hline
\end{tabular}
\label{tab1}
\end{table*}

The above optimal mode selection can serve as a theoretical bound but it is not a practical method.
We therefore devise a lightweight mode selection protocol for hybrid relaying that requires no information about the channel states and network conditions.
Recall that the task is to make a choice from two actions, \textit{active mode} and \textit{passive mode}, each with a successful transmission probability that is unknown to the relay.
Although unaware of the success probability, the relay can observe the outcome of its choice:
in every round, the relay selects a transmission mode, makes the transmission, and then receives an indication of success/failure of that transmission, or in the terminology of the bandit game, a \textit{reward} satisfying the Bernoulli distribution.
When selecting a specific mode in round $t$, the only information available to the relay is the past success transmission records of two modes up to time $t-1$.
The performance of the mode selection protocol can be measured by the \textit{regret}, which is defined as the difference between the rewards it accumulates up to time $t$ and the rewards that it would have accumulated during that period had it known from the beginning which mode had the highest expected reward.
The objective is to minimize the expected accumulated regret over all rounds.

A family of optimal-action searching policies uses the historical rewards to calculate a value called the upper confidence bound (UCB), which serves as an overestimate of the expected reward for each action in every round.
The action with the highest UCB is selected.
It has been proven that such a policy achieves sublinear regret~\cite{lattimore2018bandit}.
For the transmission mode selection problem, we adopt a specific UCB policy, called the KL-UCB (Kullback-Leibler UCB)~\cite{DBLP:journals/jmlr/GarivierC11}, which has better theoretical guarantees than the plain UCB.
Specifically, the KL-UCB for an action $i$ in round $t$ is defined as
\begin{equation}
\mbox{KL-UCB}(i, t) = \displaystyle{\mathop{\mathrm{argmax}}_{q \in [0,1]}} \;
\bigg \{ d \Big( \frac{S[i]}{N[i]}, q \Big) \leq \frac{\log (t)}{N[i]} \bigg \},
\label{kl-ucb-argmax}
\nonumber
\end{equation}
where $N[i]$ denotes the number of times that action $i$ has been selected prior to time $t$; $S[i]$ denotes the sum of the rewards obtained by choosing that action; and $d(p, q)$ is the KL divergence between the Bernoulli distribution of parameters $p, q \in [0, 1]$, which is given by
\begin{equation}
d(p, q) = p \log \frac{p}{q} + (1-p) \log \frac{1-p}{1-q}.
\label{kl-divergence}
\nonumber
\end{equation}

Like any UCB policies, the KL-UCB can manifest the exploration-exploitation in a coherent way without explicitly distinguishing the exploration/exploitation phase. 
An inspection of the KL-UCB definition reveals this trade-off: maximizing $d(S[a] / N[a], q)$ encourages the exploitation of high reward arms, while the inequality containing $\log(t) / N[a]$ encourages the exploration of less played arms.
In addition, it has been proven that for the KL-UCB, as the number of rounds $T$ tends to infinity, the expected total reward asymptotically approaches that of playing a policy with the highest expected reward, and the regret grows with the logarithm of $T$~\cite{DBLP:journals/jmlr/GarivierC11}.

It is noteworthy that the above formulation assumes a stationary distribution of rewards, which could hardly be satisfied in reality since the channel state may undergo abrupt changes, making the success probability of either mode change in an unpredictable way.
Such a problem is somehow analogous to the outdated CSI phenomena common in cooperative networks where outdated CSI can cause suboptimal relaying and transmission deterioration~\cite{DBLP:journals/tvt/Le19}.

To cope with the nonstationary environment, we adopt the idea of the Discounted UCB, in which the rewards are weighted so that recent outcomes are emphasized when calculating the expected rewards~\cite{DBLP:conf/alt/GarivierM11}.
Specifically, a discount factor $\gamma \in (0,1)$ is employed to calculate the discounted reward mean as
\begin{equation}
\bar{X}_t(\gamma,i) = \frac{\sum_{s=1}^{t} \gamma^{t-s} X_s(i) \mathbb{I}_{\{I_s=i\}}}
                           {\sum_{s=1}^{t} \gamma^{t-s} \mathbb{I}_{\{I_s=i\}}},
\label{discounted-mean}
\nonumber
\end{equation}
where $X_s(i)$ denotes the reward of arm $i$ in round $t$, and $\mathbb{I}_{\{I_s=i\}}$ is an indicator function that arm $i$ has been selected in round $s$.
The Discounted KL-UCB is then calculated as
\begin{equation}
\mbox{D-KL-UCB}(i, t) = \displaystyle{\mathop{\mathrm{argmax}}_{q \in [0,1]}} \;
\bigg \{ d \Big( \bar{X}_t(\gamma,i), q \Big) \leq \frac{\log (t)}{N[i]} \bigg \}.
\label{discounted-kl-ucb-argmax}
\nonumber
\end{equation}

The proposed Discounted KL-UCB for mode selection is outlined in Algorithm~\ref{alg1}.
It can be seen that the proposed method makes an instantaneous decision and then updates the counters sequentially based solely on the historical performance of two modes.
Such an online nature makes it adapt well to the characteristics of a hybrid relay system, where quick decisions should be made with limited prior information about the fast-changing channel states and network conditions.
\begin{algorithm}[t]
\small
\caption{Mode Selection Bandit via the Discounted KL-UCB}
\label{alg1}
\begin{algorithmic}[0]
\STATE {\bf Input}: number of transmission rounds $n$, discount factor $\gamma \in (0, 1)$
\FOR {$t=1, \ldots, 2$}
\STATE Transmit at a mode $i$ with index $i=t$
\STATE Receive an indication of success/failure as $r \in \{0, 1\}$
\STATE Initialize the select-counter of mode $i$ as {$N[i] \leftarrow 1$}
\STATE Initialize the reward-counter of mode $i$ as {$X_t(i) \leftarrow r$}
\ENDFOR
\FOR {$t=3, \ldots, n$}
\STATE Select a mode as {$i \leftarrow \mathop{\mathrm{argmax}}_{i \in \{1,2\}} \mbox{D-KL-UCB}(i,t)$}
\STATE Transmit at the selected mode $i$
\STATE Receive an indication of success/failure as $r \in \{0, 1\}$
\STATE Update the select-counter of mode $i$ as {$N[i] \leftarrow N[i] + 1$}
\STATE Update the reward-counter of mode $i$ as {$X_t(i) \leftarrow r$}
\ENDFOR   
\end{algorithmic}
\end{algorithm}

\section{Numerical Results}
\label{sec5}
\begin{figure}[t] 
\centering
\begin{minipage}[c]{0.42 \textwidth}
\includegraphics[width=1.0\textwidth]{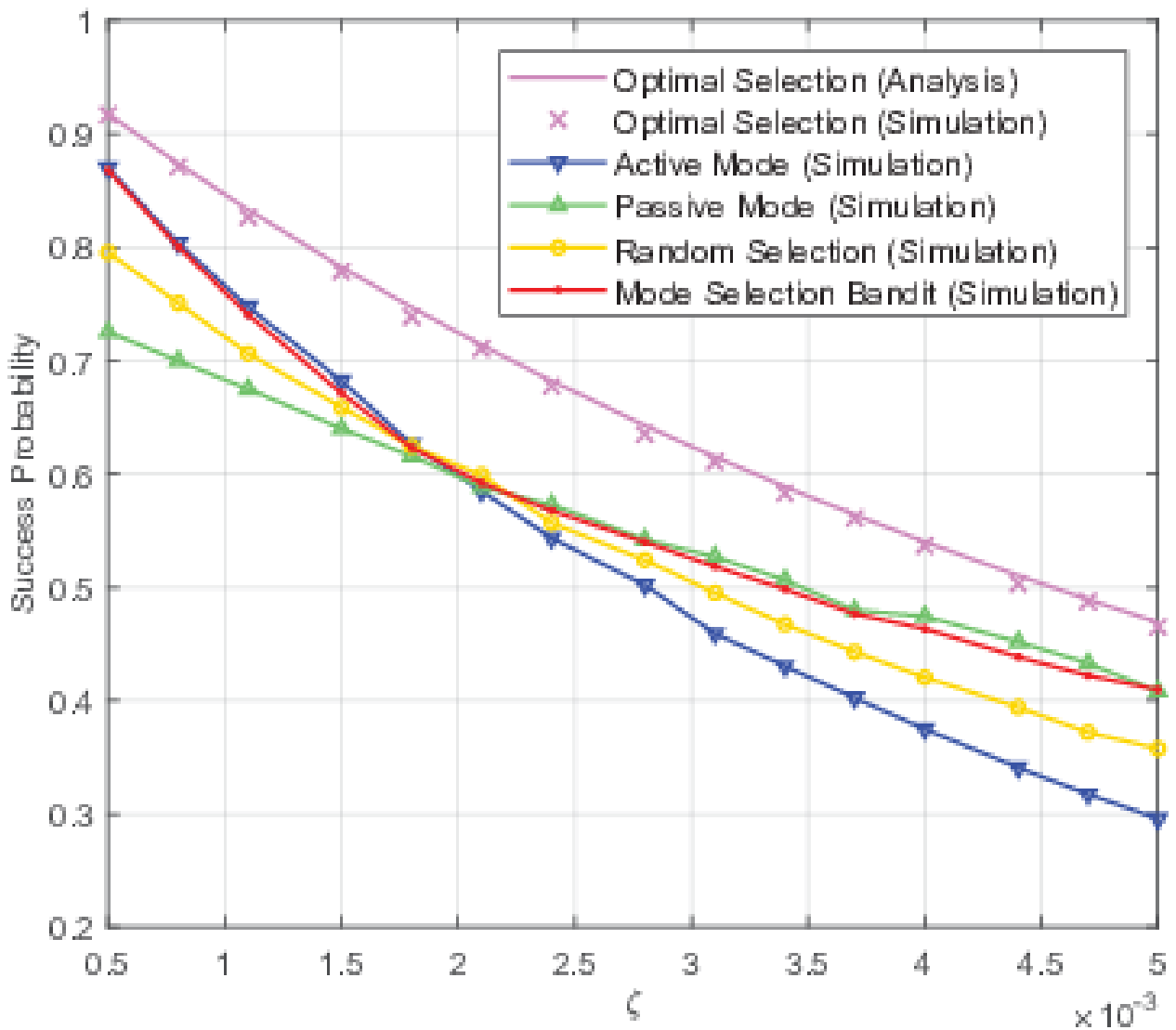}  \vspace{-7mm}
\caption{Success probability as a function of $\zeta$}  \label{fig:2}
\end{minipage}
\begin{minipage}[c]{0.42 \textwidth} 
\vspace{3mm}
\includegraphics[width=1.0\textwidth]{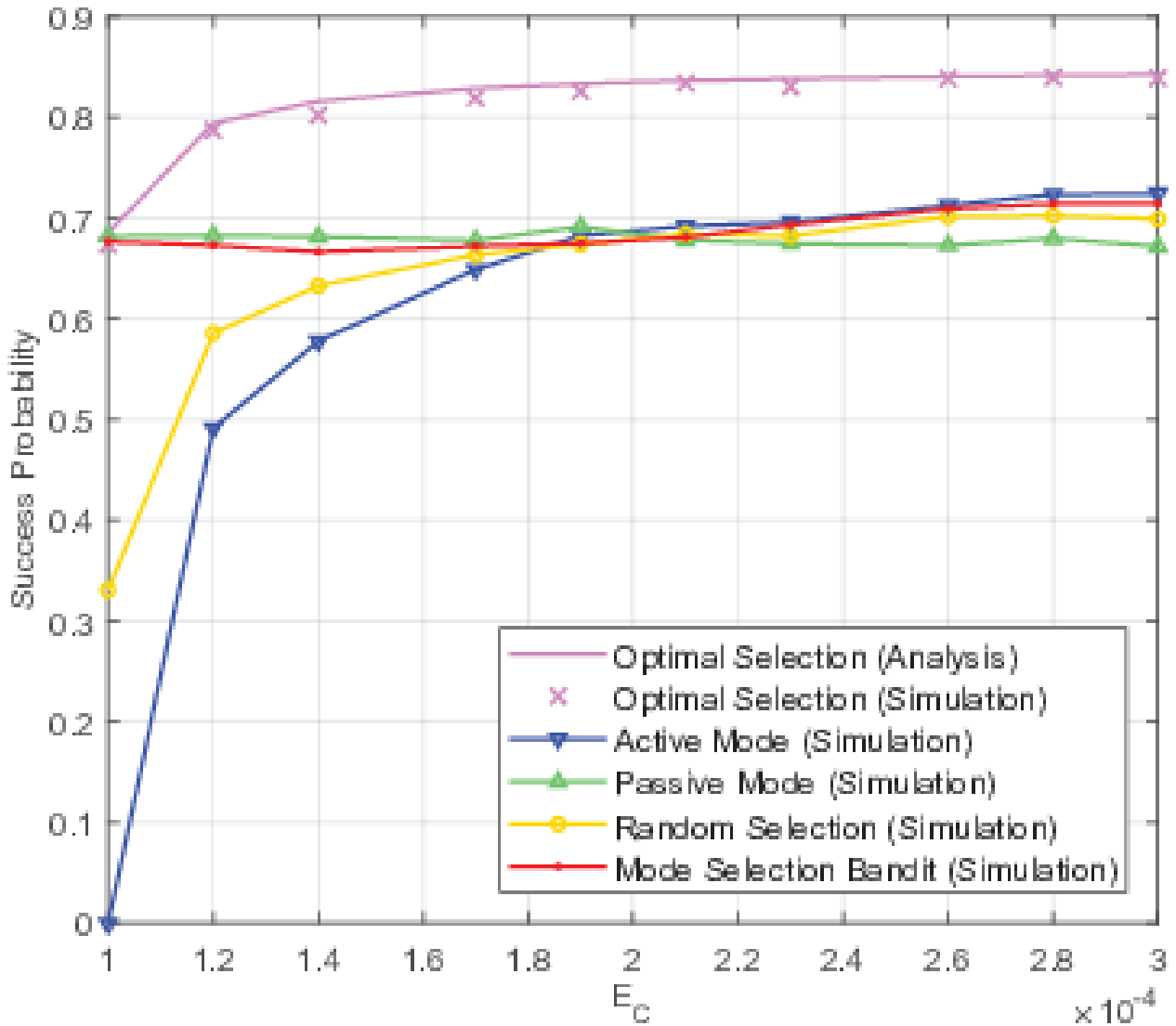}   \vspace{-7mm}
\caption{\vspace{-2mm}Success probability as a function of $E_C$} \label{fig:3}
\end{minipage}
\begin{minipage}[c]{0.42 \textwidth} 
\vspace{3mm}
\includegraphics[width=1.0\textwidth]{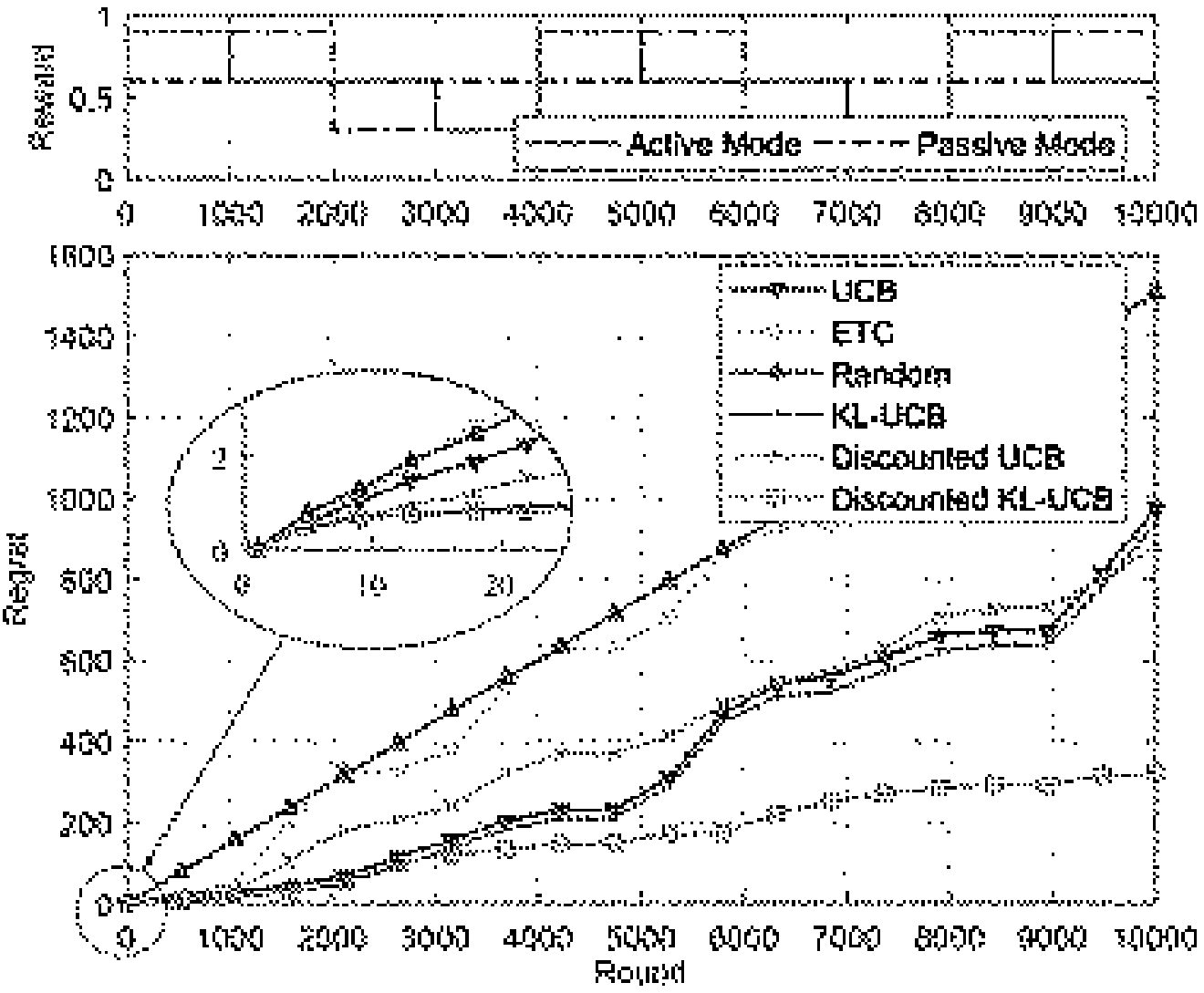} \vspace{-7mm}
\caption{Evolution of the reward distribution and regret} \label{fig:4}
\end{minipage} 
\vspace{-6mm}
\end{figure}

We perform a numerical study by simulating the source signal using a PPP with the parameters listed in Table~\ref{tab1}.
The comparison methods include Active Mode, Passive Mode, Optimal Selection (Section~\ref{sec3}), Random Selection (switching between two modes randomly), and the proposed Mode Selection Bandit (Section~\ref{sec4}).

We first change $\zeta$ and $E_C$ and examine their impacts on the success probability.
As shown in Figure~\ref{fig:2} and Figure~\ref{fig:3}, the analytical results of Optimal Selection closely match its simulation results, giving clear support for the validity of Theorem~\ref{theorem-opt}.
By comparing the success probability of Active Mode and Passive Mode, we observe that their values fluctuate with the variation in $\zeta$ and $E_C$.
This makes sticking to neither mode a good idea since neither can always win.
In contrast, by switching between the two modes dynamically, the proposed Mode Selection Bandit can approach the best performer in any occasion.

To demonstrate the effectiveness of the Discounted KL-UCB for mode selection in a varying environment, we conduct a simulation with $10,000$ rounds where the success probabilities of active mode and passive mode are two independent Bernoulli random variables with means changing every $1,000$ rounds.
Therefore, the reward distributions have changed ten times, as illustrated in the top half of Figure~\ref{fig:4}.
The comparison methods include the following:
\begin{itemize}
\item UCB: Transmission mode is selected by the classical UCB policy~\cite{DBLP:journals/wcl/MaghsudiN17};
\item ETC: Explore-then-commit policy explores both modes for a specific number of rounds (set as $100$ in here) and then sticks to the best afterwards~\cite{lattimore2018bandit};
\item Random: Transmission mode is selected at random;
\item KL-UCB: Basic KL-UCB without the discounted factor~\cite{DBLP:journals/jmlr/GarivierC11};
\item Discounted UCB: A discounted version of the UCB~\cite{DBLP:conf/alt/GarivierM11}.
\end{itemize}

Since reducing the regret is equivalent to raising the reward, or increasing the success probability for a hybrid relay, we can conclude from the cumulative regret curves in Figure~\ref{fig:4} that the performance of the Discounted KL-UCB is better than those of the others.

Note that there are fluctuations in the regret curves of all except for the proposed Discounted KL-UCB.
This is to be expected since the distributions of the rewards underwent abrupt changes ten times, and any method relying too much on outdated experiences will waste too many tries on the no-longer-optimal mode, making the regret increase in a short time.
The Discounted KL-UCB, on the contrary, reacts very fast to the distribution breakpoint, and thus enjoys a rather flat regret curve.
By inspecting the plots of the first $20$ rounds, we can observe that the Discounted KL-UCB can quickly concentrate on the optimal mode since its curve does not steadily rise.
Therefore, we conclude that the Discounted KL-UCB is better choice than existing bandit based methods when applied to select the transmission mode in an abruptly changing environment.
\section{Conclusion}
\label{sec6}

We studied the mode selection problem for a hybrid relay that can forward data through wireless-powered transmission (active mode) or ambient backscattering (passive mode).
We first derived a tractable analytical expression to characterize the end-to-end success probability of the relay with the optimal mode selection policy, which serves as an upper bound of the system's coverage performance.
We then proposed a novel bandit algorithm that adapts well to a varying environment and applied it to the transmission mode selection task.
Due to its online nature and the merit of requiring no channel state information or network conditions, the proposed bandit-based mode selection is particularly suitable for real-world systems with energy constraints. A promising future direction is to design bandit mode selection approaches for  intelligent reconfigurable surface \cite{gong2019towards,lu2019intelligent} to assist relaying. Our system model can also be extended to case with multiple hybrid relays which can integrate distributed online learning \cite{li2019data,li2019detecting} with bandit for mode selection.


\ifCLASSOPTIONcaptionsoff
  \newpage
\fi

\bibliographystyle{IEEEtran}
\bibliography{mybibliography}


\end{document}